# ALLELOPATHIC EFFECTS OF *RUMEX AZORICUS* ON LETTUCE: IMPACTS ON SEED GERMINATION AND EARLY GROWTH

Abdulrahman Ibrahim[1,2], Mariana Casari Parreira[3], Aram Akram Mohammed[4], Hawar Halshoy[5]

**ABSTRACT** – Members of the *Rumex* genus possess allelochemical compounds that vary depending on the plant part and extract concentrations. Therefore, this study aimed to investigate the allelopathic effects of extracts from the roots, stems, and leaves of *Rumex azoricus* at concentrations of 0%, 25%, 50%, and 100% on the seed germination of a lettuce plant in a laboratory setting. The results indicated that stem extract was most effective for enhancing germination percentage (68.67%), germination speed (5.12 seeds/time interval), and subsequent traits related to germination percentage (50%), germination speed (3.6 seeds/time interval), as well as subsequent traits in control seeds. The 25% extract concentration improved germination percentage (68%) and germination speed (5.05 seeds/time interval), along with subsequent traits compared to control (0%), which exhibited the lowest germination percentage (50%), germination speed (3.6 seeds/time interval), and related traits. The combined results also demonstrated that 25% stem extract significantly increased germination percentage (80%), speed (5.85 seeds/time interval), root length (1.2 cm), root fresh weight (0.032 mg), shoot length (2.2 cm), and shoot fresh weight (0.06 mg) in contrast to control seeds, which showed the minimum germination percentage (50%), speed (3.6 seeds/time interval), root length (0.17 cm), root fresh weight (0.006 mg), shoot length (0.95 cm), and shoot fresh weight (0.03 mg). The allelopathic effects of *R. azoricus* extract varied depending on the plant part and concentration; both stem and leaf extracts at low concentrations were the most effective, whereas root extracts at all concentrations produced results similar to those of control seeds.

# EFEITOS ALELOPÁTICOS DE *RUMEX AZORICUS* EM ALFACE: IMPACTOS NA GERMINAÇÃO DE SEMENTES E NO CRESCIMENTO INICIAL

**RESUMO** – Membros do gênero Rumex possuem compostos aleoquímicos que variam dependendo da parte da planta e das concentrações do extrato. Portanto, este estudo teve como objetivo investigar os efeitos alelopáticos de extratos de raízes, caules e folhas de Rumex azoricus em concentrações de 0%, 25%, 50% e 100% na germinação de sementes de uma planta de alface em um ambiente de laboratório. Os resultados indicaram que o extrato do caule foi mais eficaz para aumentar a porcentagem de germinação (68,67%), a velocidade de germinação (5,12 sementes/intervalo de tempo) e características subsequentes relacionadas à porcentagem de germinação (50%), velocidade de germinação (3,6 sementes/intervalo de tempo), bem como características subsequentes em sementes de controle. A concentração de extrato de 25% melhorou a porcentagem de germinação (68%) e a velocidade de germinação (5,05 sementes/intervalo de tempo), juntamente com características subsequentes em comparação ao controle (0%), que exibiu a menor porcentagem de germinação (50%), velocidade de germinação (3,6 sementes/intervalo de tempo) e características relacionadas. Os resultados combinados também demonstraram que 25% de extrato de caule aumentou significativamente a porcentagem de germinação (80%), velocidade (5,85 sementes/intervalo de tempo), comprimento da raiz (1,2 cm), peso fresco da raiz (0,032 mg), comprimento do broto (2,2 cm) e peso fresco do broto (0,06 mg) em contraste com as sementes de controle, que apresentaram a porcentagem mínima de germinação (50%), velocidade (3,6 sementes/intervalo de tempo), comprimento da raiz (0,17 cm), peso fresco da raiz (0,006 mg), comprimento do broto (0,95 cm) e peso fresco do broto (0,03 mg). Os efeitos alelopáticos do extrato de R. azoricus variaram dependendo da parte da planta e da concentração; tanto os extratos de caule quanto os de folhas em baixas concentrações foram os mais eficazes, enquanto os extratos de raiz em todas as concentrações produziram resultados semelhantes aos das sementes de controle.

[1] Department of Medical Laboratory Science, College of Science, University of Knowledge, Erbil, Kurdistan Region, Iraq. abdulrahman.ibrahim@knu.edu.iq, ORCID: 0000-0002-0714-6585

[2] Department of Midwifery, Erbil Technical Medical Institute, Erbil Polytechnic University, Erbil, Iraq. abdulrahman.ibrahim@epu.edu.iq,

[3] School of Agrarian and Environmental Sciences, University of the Azores, Institute for research in agricultural and environmental technology, IITAA, Terceira Island, Portugal. mariana.c.parreira@uac.pt, ORCID: 0000-0002-4939-7526

[4] Department of Horticulture, College of Agricultural Engineering Sciences, University of Sulaimani, Sulaymaniyah, Kurdistan Region, Iraq. aram.hamarashed@univsul.edu.iq, ORCID: 0000-0002-4421-3556

[5] Department of Horticulture, College of Agricultural Engineering Sciences, University of Sulaimani, Sulaymaniyah, Kurdistan Region, Iraq. hawar.hama@univsul.edu.iq, hawarhalsho@gmail.com, ORCID ID: 0000-0002-9842-6178

**Palavras-chave**: Extrato, parte da planta, concentração, velocidade de germinação

## INTRODUCTION

The genus *Rumex*, part of the Polygonaceae family and commonly known as docks and sorrels, embraces 250 species worldwide (Rao et al., 2011). *Rumex azoricus* Rech.f. is one such species native to the Azores archipelago in Portugal and is an herbaceous perennial (Dias et al., 2005). Most members of the *Rumex* genus contain allelochemical agents; however, *R. azoricus* has not been studied for its allelopathic effects. Allelopathy is a phenomenon in which one organism produces or releases biochemicals that have stimulatory or inhibitory effects on other organisms (Gam et al., 2024). Chemicals with allelopathic effects on plants include phenols, alkaloids, terpenes, non-proteinaceous amino acids, and sugars or glycosides (Korres et al., 2019). The stimulatory or inhibitory influence of allelochemical constituents depends on their concentration. At specific levels, a substance that inhibits a plant process at higher concentrations may act as a stimulator at lower concentrations (Willis, 2007). Moreover, allelopathic chemicals can be found in almost all plants and various tissues, including roots, stems, leaves, buds, flowers, and fruits (Ahmad et al., 2020; de Andrade et al., 2024). In this context, Pilipapivičius et al. (2012) applied extracts from the dry root, seeds, and shoots of *Rumex crispus* L. to spring barley seeds at different concentrations in Petri dishes. They reported that at low concentrations (0.01 to 0.1 mg), germination was enhanced, while higher concentrations (0.25 to 0.5 mg) inhibited germination. They also noted that the extract from the shoot was more inhibitory than the root extract.

Simultaneous and rapid germination is essential in agriculture for improving crop yield and quality (Wei et al., 2024). Several challenges can affect the germination rate and speed of lettuce seeds due to their requirement for specific environmental factors (Ahmed et al., 2024). Lettuce seeds are sensitive to temperature and germinate at lower temperatures depending on the cultivar (Lafta & Mou, 2013). Additionally, lettuce seed germination requires light; therefore, they should not be sown too deeply. Exogenous treatments have been applied to lettuce seeds in dark conditions to satisfy their light requirements (Gupta et al., 2019). Some treatments have been investigated for their allelopathic effects on lettuce seeds (Wang et al., 2022). Han et al. (2013) exposed lettuce seeds to different ratios of decomposed garlic stalks, and the results indicated that lettuce growth increased with low ratios of decomposed garlic stalks, while higher ratios inhibited growth. In light of the aforementioned facts, the current study aims to investigate the allelopathic effect of *R. azoricus* on the seed germination of lettuce (*Lactuca sativa*) plants.

## MATERIALS AND METHODS

This study was conducted at the Laboratory School of Agrarian and Environmental Sciences, University of the Azores, Terceira Island, Portugal, to examine the effect of different concentrations of *Rumex azoricus* extracts prepared from the root, stem, and leaves on lettuce seed germination.

**Plant collection and seed treatment**

The plant was collected from the fields of Terceira Island in Portugal. The root, stem, and leaves were then separated and chopped into small pieces, blended separately in a blender, and finally strained to separate the pulp from the extract (Figure 1). Four concentrations of the extract were prepared (0%, 25%, 50%, and 100%). The lettuce seeds were set up by placing 10 seeds on sterilized filter paper in sterilized Petri dishes, after which 2 mL of the extracts were added to the Petri dishes. The Petri dishes were then placed in an incubator (Figure 2). The seeds were checked daily to record germination speed.

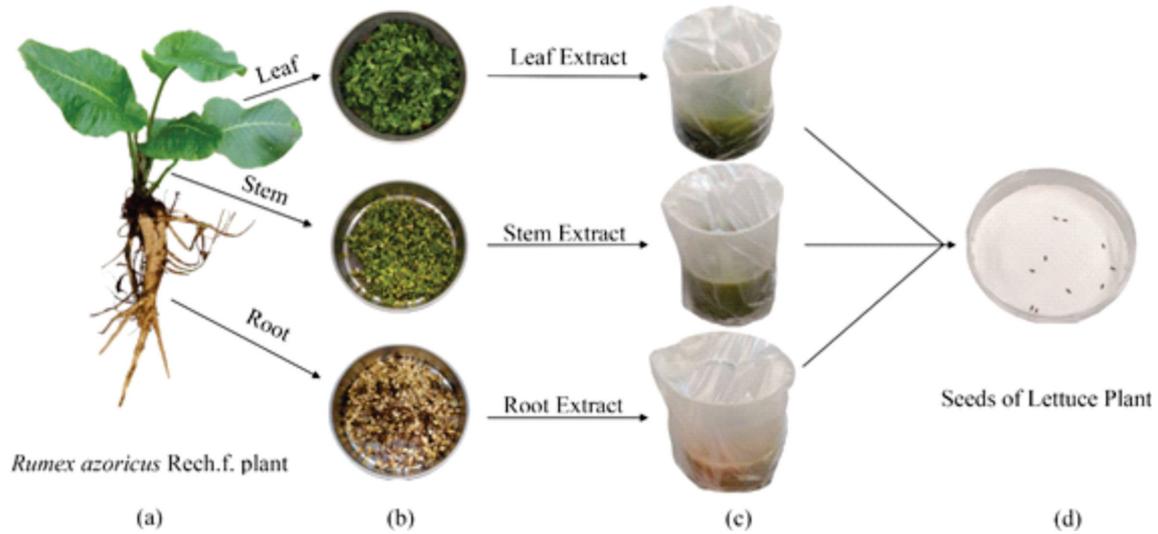

Figure 1 - Schematic illustration of extracting substances from (a) the *Rumex azoricus* plant, including its (b) root, stem, and leaves, by blending them in a (c) blender and then applying the mixture to (d) lettuce seeds in Petri dishes.

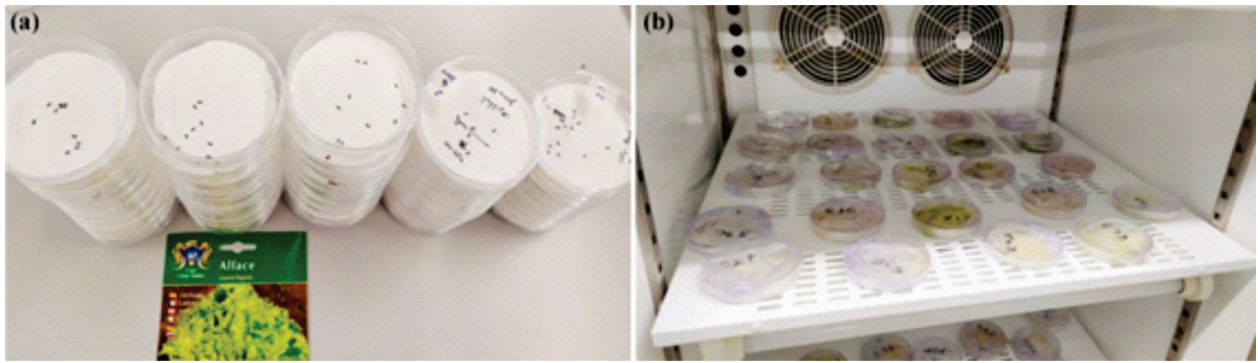

Figure 2 - Preparation of the (a) seeds on sterilized filter paper in Petri dishes and (b) placing them in an incubator.

**Statistical analysis**

        The factorial experiment was designed according to a completely randomized design (CRD) with five replications. The first factor consisted of plant part extracts (root, stem, and leaf), while the second factor represented the concentration of the extracts (0%, 25%, 50%, and 100%). A total of fifty petri dishes were used. After 10 days, germination percentage, germination speed, root length, root fresh weight, shoot length, and shoot fresh weight were assessed. The collected data were analyzed using the XLSTAT software program and Duncan's new multiple range test ($p \leq 0.05$) for mean comparison.

**RESULTS**

        In this laboratory study, extracts from different parts of *R. azoricus* were applied to lettuce seeds. The data in Figure 3 shows that the germination percentage and speed (Figures 3a and 3b) were significantly different in lettuce seeds treated with extracts from the root, stem, and leaf of *R. azoricus*. The highest germination percentage (68.67%) and speed (5.12 seeds/time interval) were observed in the seeds treated with stem extract, compared to those treated with root extract and control seeds, which exhibited the lowest germination percentages (48.67% and 50%, respectively) and speeds (3.65 and 3.6 seeds/time interval, respectively). Additionally, extracts from the stem and leaf

significantly improved root and shoot lengths compared to control seeds and those treated with root extract (Figures 3c and 3e). The longest roots (0.7 and 0.78 cm) and shoots (1.7 and 1.67 cm) were recorded in the seeds treated with stem and leaf extracts, respectively. Control seeds displayed the minimum root length (0.17 cm) and shoot length (0.95 cm), while treatment with root extract reduced root length (0.24 cm) and shoot length (0.7 cm) to the levels of the control seeds. In relation to control seeds, the treatment of lettuce seeds with extracts from the root, stem, and leaf of *R. azoricus* did not produce significant differences in the root fresh weight of the germinated seeds (Figure 3d). Regarding shoot fresh weight, there were no differences between the germinated seeds subjected to extracts from the root, stem, and leaf compared to control seeds (Figure 3f). However, the fresh weight of the shoot decreased (0.02 mg) in the seeds treated with root extract compared to those treated with stem and leaf extracts, which both had the maximum shoot fresh weights (0.05 mg).

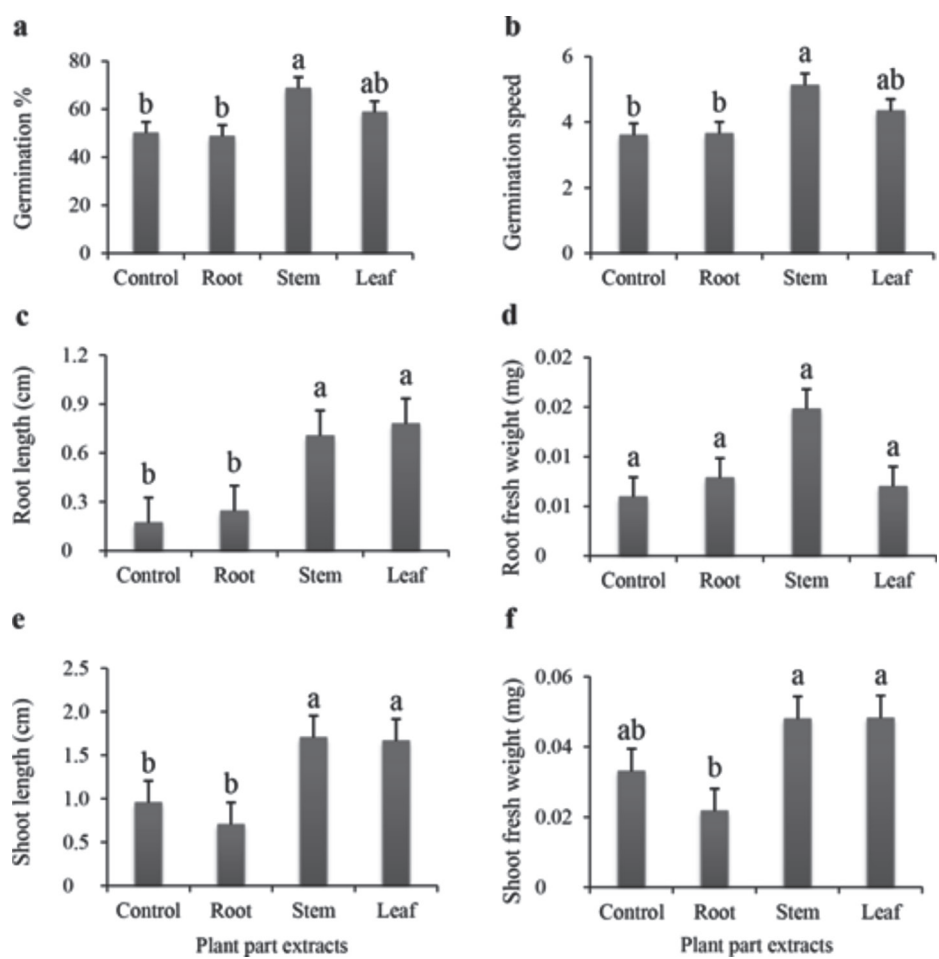

Figure 3 - Effect of *R. azoricus* root, stem, and leaf extracts on (a) germination percentage, (b) germination speed, (c) root length, (d) root fresh weight, (e) shoot length, and (f) shoot fresh weight of lettuce seeds. Bars represent means ± standard error. Columns sharing the same letter in the same figure for a single trait indicate insignificant differences according to Duncan's new multiple range test ($P \leq 0.05$).

The analyzed results of the effects of various concentrations of *R. azoricus* extracts, taken from different parts, are shown in Figure 4. The lowest concentration (25%) significantly differed in seed germination percentage and speed compared to the 0% and 100% concentrations of the extracts (Figures 4a and 4b). The 25% extracts produced the highest germination percentage (68%) and speed (5.05 seeds/time interval). In contrast, the lowest germination percentages (50% and 50.1%) and speeds (3.6 and 3.8 seeds/time interval) were recorded in the control seeds and those exposed to the 100% extract, respectively. Moreover, the concentration of extracts had a significant impact on root length and root fresh weight (Figures 4c and 4d). The 25% extracts resulted in the longest root (0.92 cm) and the heaviest root fresh weight (0.02 mg), showing a significant difference compared to the control seeds and those treated with 50% and 100% of the extracts. The shortest roots (0.17 and 0.3 cm) were observed in the control seeds and those treated with the 100% extracts, respectively. Similar lowest root fresh weights (0.006 mg) were noted in the control seeds and the seeds treated with 50% and 100% of the extracts.

Additionally, varying shoot lengths were observed due to the application of different extract concentrations (Figure 4e). The 25% and 50% extract concentrations resulted in peak shoot lengths (1.67 and 1.39 cm, respectively), while the shortest shoots (0.95 and 1.01 cm) were found in the control seeds and the seeds exposed to the 100% extracts. The seeds exposed to various extract concentrations did not show a statistically significant difference from the control seeds in terms of shoot fresh weight (Figure 4f). However, the 25% and 50% extract concentrations were significantly different from the 100% extracts regarding shoot fresh weight. The highest shoot fresh weight (0.05 mg) was recorded at 25%, while the lowest shoot fresh weight (0.03 mg) was at 100% of the extracts.

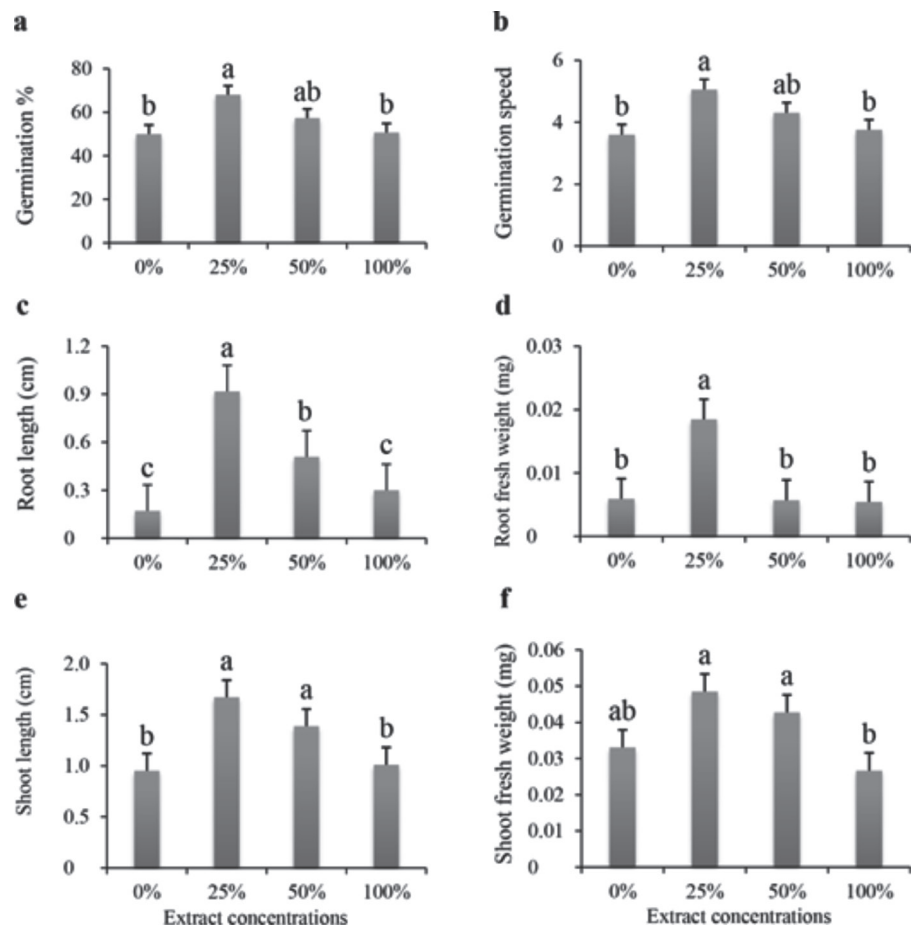

Figure 4 - Impact of varying concentrations of *R. azoricus* extracts on (a) germination percentage, (b) germination speed, (c) root length, d) root fresh weight, (e) shoot length, and (f) shoot fresh weight of lettuce seeds. Bars represent means ± standard error. Columns sharing the same letter in a single figure for a specific trait indicate insignificant differences according to Duncan's new multiple range test ($P \leq 0.05$).

　　　　The interaction effect of extracts from the root, stem, and leaf of *R. azoricus* and their varying concentrations confirmed that the stem extract at 25% significantly affected germination percentages and speeds compared to control seeds, root extracts at all concentrations, and leaf extracts at 50% and 100% (Figures 5a and 5b). The highest germination percentage (80%) and speed (5.85 seeds/time interval) were observed at 25% stem extract, whereas control seeds achieved a 50% germination rate with a speed of 3.6 seeds/time interval. In contrast, the application of root extract at 100% sharply reduced both germination percentage and speed to the lowest values (38% and 2.84 seeds/time interval, respectively). Additionally, two peaks in root lengths (1.23 and 1.27 cm) were recorded at 25% for the stem and leaf extracts, respectively, showing significant differences compared to control seeds and other treated seeds with different extracts at other concentrations (Figure 5c). Germinated control seeds displayed the shortest root length (0.17 cm). The effect of 25% stem extract on root fresh weight differed from that of control seeds and others subjected to the extracts at varying concentrations (Figure 5d). The best root fresh weight (0.032 mg) was found at the 25% stem extract; however, root extract at 50% and leaf extract at 100% resulted in the lowest root fresh weight (0.004 mg). Moreover, extracts from the stem at 25% and from the leaf at 25% and 50% yielded different shoot lengths and shoot fresh weights when compared with control seeds and seeds treated with root extracts across all concentrations (Figures 5e and 5f). Seeds exposed to stem extracts at 50% and 100% and leaf extract at 100% exhibited similar shoot lengths and shoot fresh weights to control seeds. Superior shoot lengths (2.2, 2.02, and 1.87 cm) and shoot fresh weights (0.06 mg) were recorded in seeds subjected to 25% stem extract and 25% and 50% leaf extracts, respectively. The minimum shoot length (0.55 cm) and shoot fresh weight (0.02 mg) were observed in seeds treated with 100% root extract. Control seeds showed a shoot length of 0.95 cm and a fresh weight of 0.03 g.

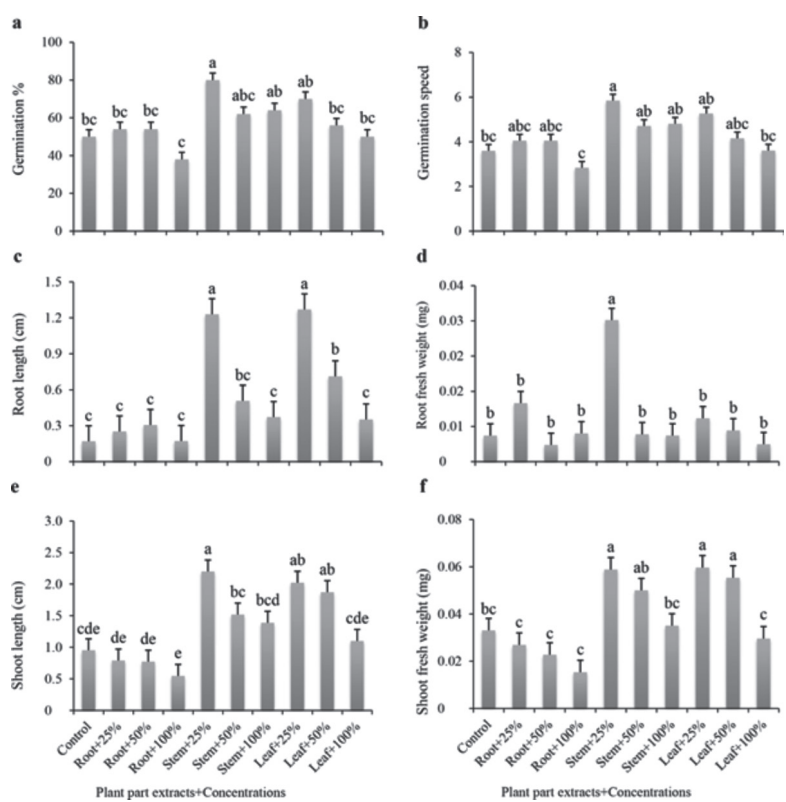

Figure 5 - Interaction effect of extracts from various plant parts of *R. azoricus* and their different concentrations on (a) germination percentage, (b) germination speed, (c) root length, (d) root fresh weight, (e) shoot length, and (f) shoot fresh weight of lettuce seeds. Bars represent means ± standard error. Columns that share the same letter in the same figure of a single trait indicate insignificant differences according to Duncan's new multiple range test ($P \leq 0.05$).

**Interrelationship of the variables**

Principal component analysis (PCA) shown in Figure 6 indicated that the variables were distributed among two PCs; the first accounted for 82.7% of the variance, while the second comprised only 10.39%. Consequently, stem extract at 25% (Stem 25%) positioned on the positive sides of both PCA1 and PCA2 strongly clustered with germination percentage (G%) and speed (GS), and Stem 25% was also the treatment nearest to root fresh weight (RFW). Leaf extract at 25% (Leaf 25%) was located on the positive side of PCA1 and grouped with root length (RL), shoot length (SL), and shoot fresh weight (SFW). Stem extracts at 50% and 100% (Stem 50% and Stem 100%) were centered in the PCA. Moreover, root extracts at 25%, 50%, and 100% (Root 25%, Root 50%, and Root 100%) were positioned on the negative side of PCA1 and close to the zero value of PCA2, showing a negative association with RL, SL, and SFW. Furthermore, the control and leaf extract at 100% (Control and Leaf 100%) were situated on the negative sides of both PCA1 and PCA2 and were negatively correlated with RFW, GS, and G%.

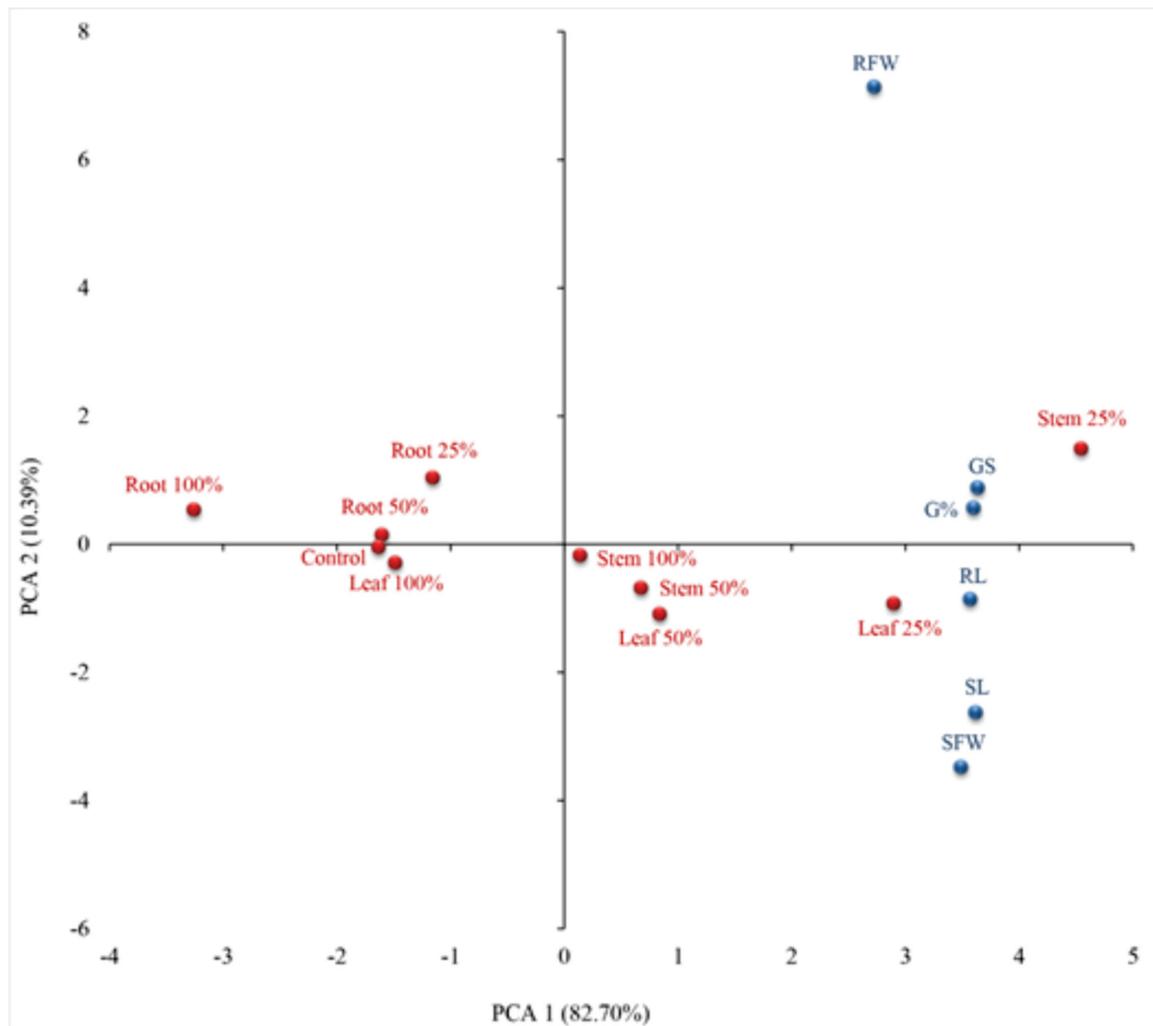

Figure 6 - The biplot from principal component analysis (PCA) illustrates the relationships among the variables, including treatments and studied parameters. G%: germination percentage, GS: germination speed, RL: root length, RFW: root fresh weight, SL: shoot length, and SFW: shoot fresh weight.

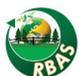

Figure 7 illustrates that G% was positively correlated with GS (r = 1, P = 0.0001), RL (r = 0.84, P = 0.002), RFW (r = 0.69, P = 0.028), SL (r = 0.86, P = 0.001), and SFW (r = 0.81, P = 0.004). Additionally, a positive correlation was observed between GS and RL (r = 0.83, P = 0.003), RFW (r = 0.65, P = 0.04), SL (r = 0.85, P = 0.002), and SFW (r = 0.81, P = 0.005). RL also showed a positive correlation with RFW (r = 0.64, P = 0.047), SL (r = 0.91, P = 0.0002), and SFW (r = 0.88, P = 0.001). Furthermore, SL and SFW were positively correlated (r = 0.97, P = 0.0001).

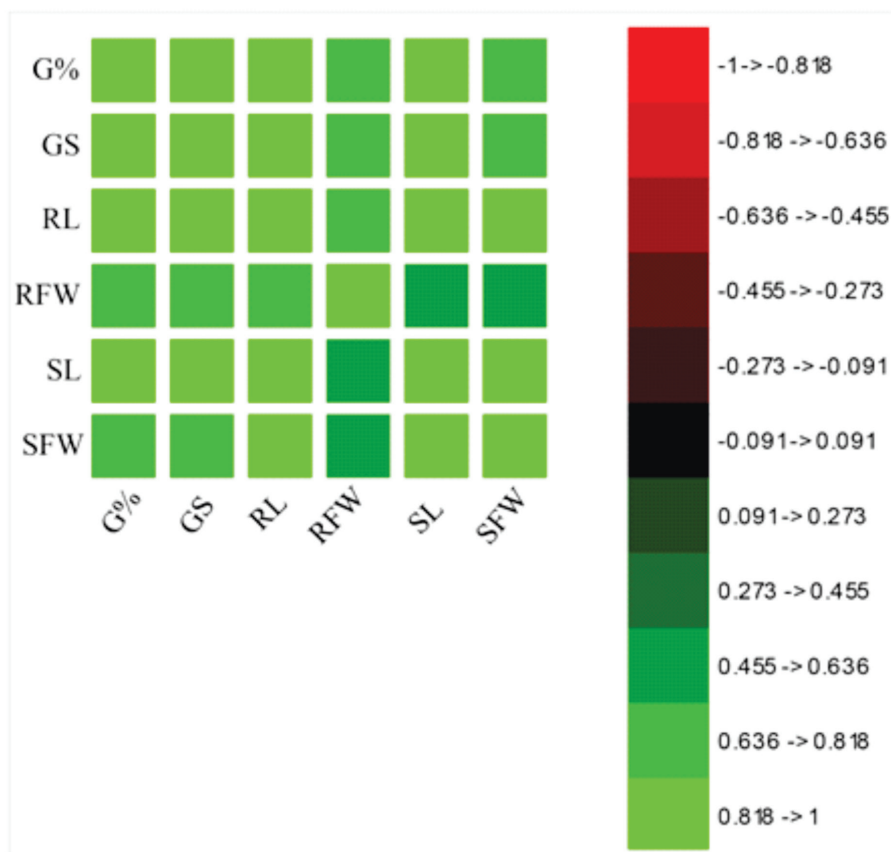

Figure 7 - Correlation ($P \leq 0.05$) among the measured parameters of lettuce seeds under the influence of extracts from various plant parts of *R. azoricus* and their different concentrations. G%: germination percentage, GS: germination speed, RL: root length, RFW: root fresh weight, SL: shoot length, and SFW: shoot fresh weight.

## DISCUSSION

In the current study, the allelopathic effects of *R. azoricus* extract from the root, stem, and leaf at 0%, 25%, 50%, and 100% were investigated. The results varied depending on the plant part from which the extracts were derived (Figure 3). Seed germination and speed increased due to the stem extract, while root and shoot lengths improved with the application of stem and leaf extracts. Notably, the root extract decreased shoot fresh weight. Overall, the root extract was ineffective, producing results similar to untreated (control) seeds. Correspondingly, Pilipapivičius et al. (2012) reported that the allelopathic effect of *Rumex crispus* varied depending on the plant part from which the extract was prepared. These variations may be attributed to the differences in chemical composition and concentration among the plant parts. The leaves and roots of *Rumex* spp. were evaluated to determine the level of secondary metabolites (oxalates, phenols, and tannins) that may exert allelopathic influences, and it was found that the roots contained higher tannin levels (Costan et al., 2023). Moreover, extracts of the root and shoot of Rumex japonicus were applied to seeds of lettuce, and more inhibition in shoot growth was noted due to the extract from the root more than from the shoot. This inhibition was ascribed to high protocatechuic, p-hydroxybenzoic, ferulic acids, and pyrocatechin in root (Elzaawely et al., 2005).

On the other hand, the concentrations tested in this study obtained various results. The 25% concentration was exceptional for germination percentage and speed, root length, and root fresh weight (Figure 4). However, as the concentration increased, the germination percentage and speed, root length, and root fresh weight decreased, reaching the control (0%) level at 100%. Both 25% and 50% concentrations of *R. azoricus* were optimal for shoot length and shoot fresh weight, whereas the 100% concentration caused a decline in shoot fresh weight. Other researchers have found that the allelopathic effects of *Rumex* spp. extracts are concentration-dependent. In a study on the extract of *R. acetosella*, it was found that the inhibitory effect of the extract increased with higher concentrations (Gam et al., 2024). This may be attributed to the increased presence of inhibitory compounds at higher concentrations. In this context, the suppression of weed growth by allelopathic plant extracts (*Dipteryx lacunifera* Ducke, *Ricinus communis* L., *Piper tuberculatum* Jacq., and *Jatropha gossypiifolia* L.) has been linked to the greater concentrations of allelochemicals present in these extracts (Lopes et al., 2022).

Combinations of various extracts from different parts (root, stem, and leaf) of *R. azoricus* and extract concentrations showed that the stem extract at 25% was the most effective treatment for all studied parameters (Figure 5). Additionally, the leaf extract at 25% for root length and at 25% and 50% for shoot fresh weight registered results similar to those of the stem extract at 25%. Root extract at all concentrations produced results comparable to the control seeds. However, the root extract at 100% reduced germination percentage and speed, shoot length, and shoot fresh weight compared to control seeds, though not to a statistically significant extent. Similarly, in a study on the allelopathic effect of *Rumex dentatus* by El-Beheiry et al. (2019), it was achieved that the shoot extract of this species was slightly raised seed germination in wheat at 20%, but at higher concentrations (40% and 50%) decreased germination; however, root extracts inhibited germination at all concentrations. Furthermore, in the current study, the PCA emphasized that the stem and leaf extracts at 25% were the most effective treatments, as they were closely associated with the studied parameters on the farthest positive sides of both PCAs for Stem 25%, and on the positive side of PCA1 for Leaf 25% (Figure 6). In contrast, control seeds, root extracts at 25%, 50%, and 100%, and leaf extract at 100% were located at the farthest negative side of PCA1 and at zero value on PCA2, indicating that they exhibited the minimum values. Moreover, stem extracts at 50% and 100%, along with leaf extract at 50%, were positioned at the center of the PCA, suggesting that they were intermediate in results and did not differ significantly from both the highest and lowest values. Furthermore, Pearson correlation analysis clarified that germination speed was positively correlated with the growth characteristics of germinated lettuce seeds (Figure 7). Previous studies have reported that rapid germination facilitates earlier root establishment and enhances water and nutrient uptake. Consequently, these factors contribute to accelerated growth of both the root and shoot, resulting in increased biomass (Mohammed et al., 2023). The figure also demonstrated a positive association between root length and root fresh weight, as well as shoot length and shoot fresh weight. These correlations may be attributed to the high root and shoot lengths, which lead to the accumulation of more cells and tissues, thereby increasing the fresh weight of both the root and shoot (Alberts et al., 2002).

## CONCLUSION

In the current study, the analyzed data indicated that the allelopathic effect of *R. azoricus* varied among different plant parts depending on concentration. The above-ground parts (stem and leaves) had a stimulatory influence on lettuce seeds at low concentrations, but at high concentrations, they were not different from the untreated control seeds. However, root extracts showed no difference from the control seeds at any concentration studied. Further research is needed to identify the allelopathic compounds that may exist in this species, and its allelopathic influence is better to be tested with the seeds of other species.

## AUTHERS' CONTRIBUTION

Abdulrahman Ibrahim and Mariana Casari Parreira designed the experiment, performed it, and collected the data. Hawar Halshoy arranged and analyzed the data, and created the figures. Aram Akram Mohammed wrote the draft of the manuscript. Finally, all authors discussed and commented on the results and contributed to the final manuscript.